\documentclass[twocolumn,showpacs,preprintnumbers,amsmath,amssymb,amsfonts]{revtex4}

\usepackage{graphicx}
\usepackage{dcolumn}
\usepackage{bm}
\usepackage{colordvi}

\begin{document}

\preprint{APS/123-QED}

\title{Velocity oscillations in confined channel flows of concentrated colloidal suspensions}

\author{Lucio Isa}
\author{Rut Besseling}%
\author{Alexander N. Morozov}%
\author{Wilson C. K. Poon}%
\affiliation{%
SUPA \& School of Physics, The University of Edinburgh, James Clerk Maxwell Building, The Kings Buildings, Mayfield Road, Edinburgh EH9 3JZ, UK}%

\date{\today}

\begin{abstract}
We study the pressure-driven flow of concentrated colloids confined in glass micro-channels at the single particle level using fast confocal microscopy. For channel to particle size ratios $a/\bar{D} \lesssim 30$, the flow rate of the suspended particles shows fluctuations. These turn into regular oscillations for higher confinements ($a/\bar{D} \simeq 20$). We present evidence to link these oscillations with the relative flow of solvent and particles (permeation) and the effect of confinement on shear thickening.

\end{abstract}



\maketitle

Confined flows are important in both fundamental and applied science. In simple liquids, confinement effects show up on the ten-molecule, or $\sim 10$~nm, lengthscale~\cite{granick,klein}. The properties of soft and granular matter, however, are controlled by mesoscopic objects (polymers, colloids,  grains), whose sizes can be $\mu$m to mm or beyond. Here, confinement effects can be relevant on length scales that are macroscopic, e.g. in the silo flow of grains~\cite{lepennec}, or at least on length scales resolvable in the optical microscope, such as the flow of suspensions of micron-sized particles in $\sim 10-100\mu$m devices. Thus, for example, confinement effects in the nozzle-directed printing of complex structures using colloidal inks~\cite{Lewis02,Lewis05} have recently been found~\cite{Lewis07}. More generally, colloids are increasingly found in microfluidic applications~\cite{Microfluidics}. 

Here we present experiments on the flow of a hard-sphere suspension at nearly random close packing, a `paste'~\cite{PasteReview}, confined in 20 to 40 particle diameter square channels. Strong confinement is found to affect the quiescent properties of dense colloids, e.g. the onset of the hard-spheres glass transition~\cite{nugent}, and therefore we may expect strong effects under flow, too. By tracking particles using fast confocal microscopy, we indeed find large-amplitude velocity oscillations in the narrowest channels. 

Oscillations are ubiquitous in paste flow~\cite{yaras,lukner,haw1}, but their origin remains unclear. The oscillations we see are clearly due to confinement --- they are prominent in 20-diameter channels, but vanish for 40-diameter channels. Interestingly, some channels used in some previous experiments, when expressed in particle diameters, are of the same order of magnitude as those used in our work: $\sim 10-40$ particle diameters in \cite{yaras} and $\sim 50-100$ diameters in \cite{lukner}; it is possible that confinement may also be relevant in these cases.

\begin{figure}[h]
\center{\includegraphics[width=0.47\textwidth,clip]{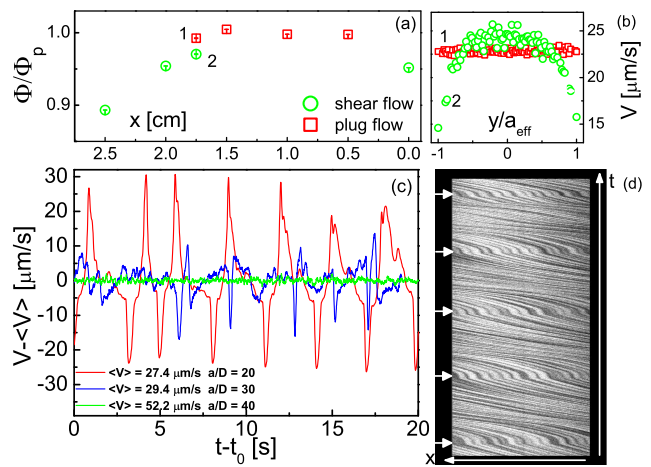}}
\vspace{-0.25cm}
\caption{\label{fig1} 
(a) Normalized particle volume fraction as a function of $x$ from particle counting in 1000 frames. (b) Velocity profiles as a function of normalized $y$ ($a_{\rm eff} = a - \bar{D}/2$) corresponding to points 1 and 2 in (a). (c) Velocity offset by the long-time average as a function of time for three different channel sizes. (d) Space-time ($x$-$t$) diagram of oscillating flow in a $\simeq 20 \bar{D}$ channel. The arrows highlight jamming events.}
\end{figure}

We used sterically-stabilized polymethylmethacrylate (PMMA) spheres (diameter $\bar{D} = 2.6 \pm 0.1$~$\mu$m; volume fraction $\Phi \gtrsim 0.63$, from confocal microscopy; Brownian time $\tau_{\rm B} = 4.4$s), fluorescently labelled with nitrobenzoxadiazole and suspended in a mixture of cyclo-heptylbromide and mixed decalin (viscosity 2.6 mPa$\cdot$s) for buoyancy matching at room temperature. A pressure difference, $\Delta P$ ($\approx \,10^2$ to $10^4$ Pa), is applied to drive the suspension into square borosilicate glass micro-channels (Vitrocom Ltd; width $a = 50$, 80 and $100 \mu$m) with untreated, smooth inner walls. The channel dimensions ($a/\bar{D} \simeq 20$, 30 and 40) were not integral multiples of $\bar{D}$ to avoid complete layering. 

The flow across the full width of the channel $y$ was imaged with a Visitech VTeye confocal scanner and a Nikkon TE 300 inverted microscope. We collected $44 \mu$m$\times 58 \mu$m images (107 frames/s) at 17 $\mu$m from the lower surface. From these we located (resolution $\simeq 50$~nm) and tracked particles~\cite{crocker,isa1}. From the single particle velocities we calculated the macroscopic flow velocity, averaged over five frames to eliminate high frequency instrumental and tracking noise. 

In the widest channel ($a/\bar{D} \simeq 40$), the velocity is steady with time, Fig.~\ref{fig1}-c~\footnote{We report the flow velocity $V$ offset by the long-time average  $\langle V \rangle$, where the latter is calculated on windows 2-3 times larger than the typical oscillation periods.}. The velocity profile and density (data not shown) do not vary with position along the channel $x$ after entrance effects. The velocity profile consists of shear zones near the walls and a central plug with a flow-speed-independent size. We have recently shown that such profiles can be understood using a model developed for granular pipe flow \cite{isa2}. 

\begin{figure}[h]
\center{\includegraphics[width=0.45\textwidth,clip]{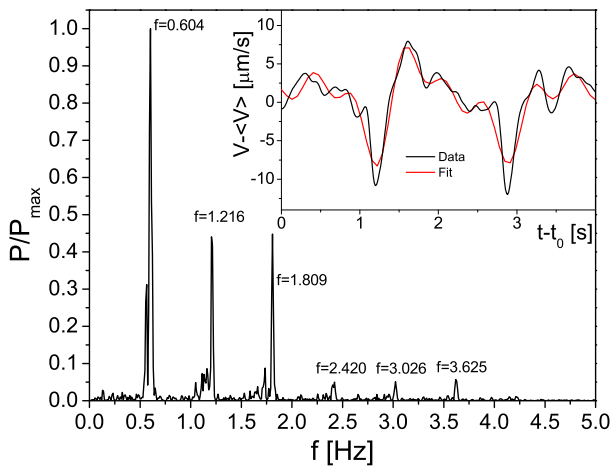}}
\caption{\label{fig3} Normalized power spectrum for the velocity signal in the inset to Fig.~\ref{fig2} ($\langle V \rangle =  20.4 \mu$m/s). Inset: four seconds of the offset velocity signal with the fit obtained from the superposition of 6 sine waves from the power spectrum (red). }
\end{figure}

The situation is radically transformed when the channel size is halved (to $a/\bar{D} \simeq 20$). The average velocity profile and density are now $x$-dependent, Fig.~\ref{fig1}-a. Near the inlet ($x=0$) we find the same kind of sheared profile as in the wider channels. Further downstream we encounter another region where the central plug extends across the full channel, i.e. shear between particles is completely absent and the suspension flows as a solid body with velocity $V$; in this \textit{complete plug} region the density reaches a maximum value $\Phi_{\rm p}$, and remains constant over $\sim 1$cm, before terminating even further downstream at a sharp interface with another sheared region, Fig.~\ref{fig1}-a, whose density decreases away from the plug. Dramatic, quasi-regular velocity oscillations are also seen, Fig.~\ref{fig1}-c. Such oscillations are nearly washed out in a $\simeq 30\bar{D}$ channel, and absent when the channel is widened to $\simeq 40\bar{D}$. 

Fig.~\ref{fig1}-d shows a `space-time diagram' of the oscillations in the $\simeq 20\bar{D}$ channel, constructed by stacking one-pixel-wide $y$-bins taken from consecutive images in the channel's center, in the region of complete plug flow. Changes in the slope of the traces correspond to changes in the flow speed, and we observe the presence of regularly spaced horizontal bands corresponding to events where the flow comes to an almost complete arrest (\textit{jams}). The power spectrum of the oscillations, Fig.~\ref{fig3}, consists of a fundamental frequency and higher harmonics; superposing the fundamental and the first five harmonics reproduces the gross features of the signal (inset). 

\begin{figure}[h]
\center{\includegraphics[width=0.45\textwidth,clip]{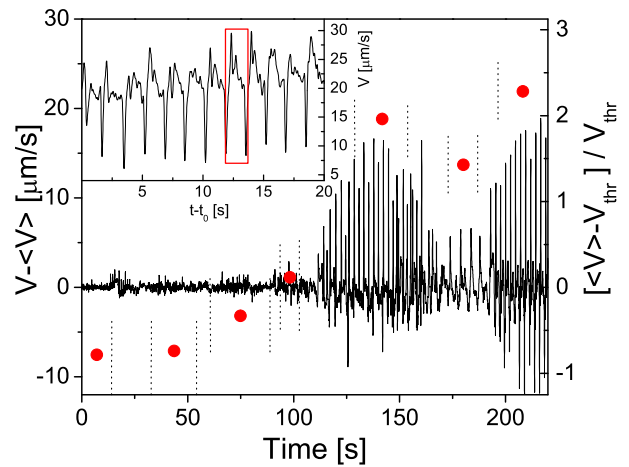}}\vspace{-0.25cm}
\caption{\label{fig2}Flow velocity $V(t)$ (black, left axis), offset by the long-time average $\langle V \rangle$. Right axis: corresponding mean $\langle V \rangle$ relative to the threshold for oscillations (normalized by $V_{\rm thr} = 7\mu$m/s, red bullets), obtained by changing the applied $\Delta P$ in the highlighted time windows (veritcal segments). Inset: example velocity oscillations, with one cycle highlighted (red).} 
\end{figure}

These regular jams are absent below a particular threshold average flow rate $V_{\rm thr}$. Fig.~\ref{fig2} shows $V - \langle V \rangle$ and the corresponding normalized distance from the threshold $[\langle V \rangle - V_{\rm thr}]/V_{\rm thr}$ versus time ($a \simeq 20 \bar{D}$). For small $\langle V \rangle$ the flow is time-independent, but above the threshold oscillations appear. Reducing the average flow rate reduces the amplitude of the velocity fluctuations.
 
\begin{figure}[h]
\center{\includegraphics[width=0.4\textwidth,clip]{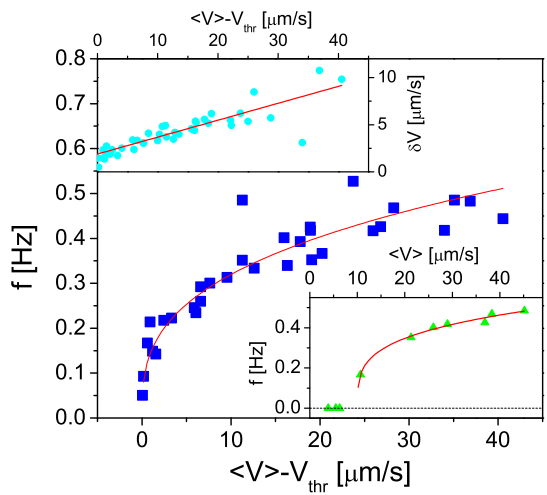}}\vspace{-0.25cm}
\caption{\label{fig4}Fundamental frequency of the velocity oscillations as a function of $\langle V \rangle - V_{\rm thr}$; the full curve is a power-law fit with exponent $0.34 \pm 0.13$. Top inset: amplitude (standard deviation) of the velocity oscillations as a function of $\langle V \rangle - V_{\rm thr}$ (points) and a linear fit (line). Bottom inset: fundamental oscillations frequency as a function of the average flow speed $\langle V \rangle$ for one run (points) with a power-law fit (curve).} 
\end{figure}

We measured the fundamental frequency of the oscillations as a function of the local average speed $\langle V \rangle$ (Fig.~\ref{fig4})~\footnote{This can also be done in terms of $\Delta P$,  but the former represents better the microscopics of the flow, since the pressure difference is only controlled macroscopically and we are not able to access its local values. Moreover the relation between applied $\Delta P$ and measured $\langle V \rangle$ can be non-monotonic and strongly history dependent.}.
The flow is steady for average velocities below a run-dependent threshold, while oscillations arise above this threshold (bottom inset, Fig.~\ref{fig4}). The frequencies $f$ can be fitted by $\displaystyle{f(\langle V \rangle) = \alpha (\langle V \rangle - V_{\rm thr})^{\beta}}$, where $f$ extrapolates to 0 at $ V_{\rm thr}$ (for the curve in the inset $ V_{\rm thr} = 10.06 \mu$m/s) and $\alpha$ and $\beta$ are fitting parameters. Plotting $f$ versus  $\langle V \rangle - V_{\rm thr}$ collapses all the data onto one single curve, which is in turns fitted by a power law with exponent $\beta = 0.34 \pm 0.13$. 

The amplitude of the oscillations (estimated by the standard deviation of the velocity signal, Fig.~\ref{fig4}, top inset) shows a linear increase with flow speed above the threshold. The degree to which the linear fit misses the origin represents the upper limit of the noise in non-oscillating signals. This linear dependence is directly related to the fact that at each jamming event, the velocity drops from roughly $\approx \langle V \rangle$ to a much lower value (see the `dips' in the inset to Fig.~\ref{fig2}) that is only weakly dependent on the average flow speed.

Summarizing our observations, we find that under strong confinement ($a/\bar{D} \simeq 20$) and above a threshold flow rate, the flow velocity oscillates while driven by a constant pressure difference. The frequency and the amplitude of the oscillations increase with flow rate. The velocity profile and particle density are non-uniform along the channel. The velocity oscillations and spatial non-uniformities along the flow directions disappear when the channel width is doubled ($a/\bar{D} \simeq 40$).  

\begin{figure}[htb]
\center{\includegraphics[width=0.45\textwidth,clip]{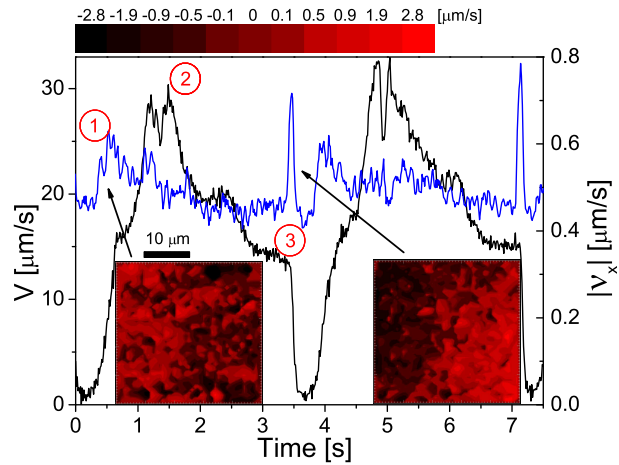}}
\caption{\label{fig5} Plug velocity as a function of time in the lab frame (right axis) and absolute value of the $x$-component of particle velocities in the comoving reference frame. The values of $\nu_{\rm x}$ have been calculated using particle displacements over 0.1 s. The inset diagrams show the spatial distribution of $\nu_{\rm x}$ over the field of view, with flow direction being from right to left.}
\end{figure}

To understand the physical origins of these oscillations, we analyse a typical cycle of the velocity of the plug region in detail (black curve, Fig.~\ref{fig5}). For convenience, we start at a time when the plug is jammed (due to a mechanism that we will shortly propose), Point 1 in Fig.~\ref{fig5}. 

The (constant) pressure gradient along the channel will drive {\it permeation} flow of the solvent relative to the plug. We suggest that this erodes particles from the front of the plug, and gives rise to a less dense region; such a region is indeed observed, Fig.~\ref{fig1}-a. A density wave should therefore propagate backwards, slightly expanding the plug, and allowing the whole plug to accelerate from rest, Point~2, Fig.~\ref{fig5}. We see evidence for such a back-propagating dilatorial wave by analysing particle displacements in the plug's co-moving frame. The blue curve in Fig.~\ref{fig5} reports the absolute value of the $x$-component of the particle velocity in this frame, $|\nu_{\rm x}|$. We observe a rise as the plug unjams. The (signed) values of this velocity are represented in the left-hand inset image, and clearly show an expansion of the plug. 

The less dense region at the front of the plug can sustain shear (point 2, Fig.~\ref{fig1}-a). We estimate that the shear rates here (in terms of the Pecl\'{e}t number,~Pe) are in the region $2 \lesssim {\mbox Pe} \lesssim 20$. In this regime, concentrated suspensions are expected to shear thicken~\cite{frith}: a consequence of the inability of particles to move past each other quickly at high particle densities. We suggest that as the shear front begins to thicken and slows down, particles accumulate at its back (upstream), increasing the local density and enhancing the tendency to shear thicken, until the suspension suddenly jams. 

In this scenario, we expect a compression wave passing through the system at this point. The blue curve in Fig.~\ref{fig5} does indeed show a sharp peak at Point 3. The (signed) values of this velocity represented in the right-hand inset also show this compression. The cycle now repeats.

The existence of a threshold flow rate (Pe~$\simeq 2$) for oscillations can be traced back to a critical rate needed for the onset of shear thickening. Moreover as recently found by Fall et al.~\cite{fall}, confinement plays a crucial role in the occurrence of shear thickening, with the critical shear rate for shear thickening decreasing as the gap size decreases in the 10-100 particle diameter regime. Apparently, then, within the range of flow rates we can explore, the channel needs to be smaller than $\simeq 30\bar{D}$ for shear thickening, and therefore oscillations, to occur.

The interplay between permeation and shear thickening can therefore give rise to oscillations in narrow channels. However, the {\em regularity} of the oscillations observed under strong confinement requires further discussion.

\begin{figure}[htb]
\center{\includegraphics[width=0.42\textwidth,clip]{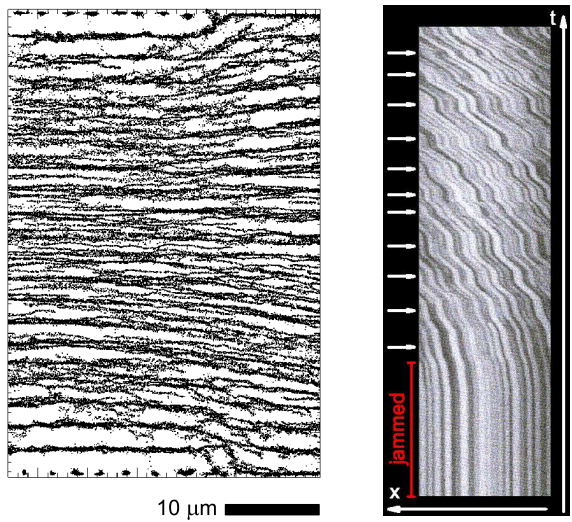}}
\caption{\label{fig6}Left: superposition of particles positions during flow (right to left, $\langle V \rangle \simeq 25 \mu$m/s, 50 images) at the plug-shear interface showing converging streamlines. Right: space-time diagram of erratic flow in a $\simeq 20 \bar{D}$ channel with rough walls. The arrows highlight jamming events.}
\end{figure}

We have already suggested that permeation unjams the system by eroding particles at the plug-shear interface. Given that particles close to the centerline move faster in the shear region, (Fig.~\ref{fig1}-b), the streamlines in the plug-to-shear transition zone must be \textit{convergent} (Fig.~\ref{fig6}, left), reminiscent of particle movement in the vicinity of an imposed geometrical constriction, such as an hourglass or funnel. There is, however, an important difference. The colloids in our case \textit{self-organize} into a `constriction with convergent flow' scenario, i.e. a fraction of the colloids consitute the constriction into which the rest of the particles flows convergently. We suggest that it is such self-organization, with appropriate feedback, that creates the regular oscillations we see under high confinement. 

If our surmise is correct, then we expect the oscillations to become much less regular if the geometric constriction is imposed rather than self-organized. This is indeed seen if the channel walls are made rough on the particle scale (by sintering a disordered layer of colloids onto the channel walls~\cite{isa2}). Boundary roughness hinders the presence of a complete plug and imposes jamming at specific (`quenched') locations in the channel. Fig.~\ref{fig6} (right) shows a space-time diagram of a flow in a $\simeq 20 \bar{D}$ channel with rough walls. Under an intial applied $\Delta P$ the flow jams completly at time $t_0$ (not in the figure) and stays jammed until $\Delta P$, and thus permeation, is increased tenfold. The flow then starts again with a series of {\em erratic} jamming events. Erratic jamming was also reported by Haw in the proximity of a fixed constrction in the flow of sterically-stabilised PMMA colloids~\cite{haw1}.

The mechanism described above is suggestive of similarities with the case of dry grains flowing in an hourglass \cite{wu}, where the regularity is due to the interplay of air counterflow and jamming of the particles at the constriction \cite{lepennec2,muite}. Here, the regularity is inherent to the particle dynamics under confined shear flow. The `self-organized' constriction and jamming in our channels appear only if the shear rates are sufficiently high, i.e. if the suspension is shear thickening. Moreover, for rates above the shear thickening threshold, we measured that it takes a certain accumulated strain ($\sim 2$ to 10) for the system to jam, suggesting analogies with strain hardening in colloidal suspensions~\cite{melrose2}. The strong confinement effects we observe also recall analogies in dry granular flow ~\cite{tsai,horluck1}. 

In conclusion, we have observed strong oscillations in the flow of concentrated hard-sphere-like colloids in square channels. Oscillations are widespread in the bulk flow of pastes. In previous studies using polydisperse industrial particles, oscillations were ascribed to filtration~\cite{yaras} and non-specific `stick-slip' instabilities~\cite{lukner}. By studying a well-characterised model suspension using real-time, single-particle imaging, we have been able to propose a specific, microscopic mechanism for the onset of instability: an interplay between solvent permeation and shear thickening in the plug-to-shear transition zone. We observe such instabilities only in channels narrower than $\sim 30 \bar{D}$, likely because of confinement effects on shear thickening \cite{fall}. Interestingly, confinement effects are probably also relevant in previous studies ($\sim 10-40 \bar{D}$ \cite{yaras} and $\sim 50-100 \bar{D}$ \cite{lukner}, but note the much higher Pe because of particle size in both cases). Irrespective of whether our proposed mechanism is also relevant for these industrial particulates, our results should have direct implications for the flow of concentrated colloids in microfluidic devices \cite{Lewis07,Microfluidics}. 

We thank A.B. Schofield for particles and M.E. Cates for illuminating discussions. LI was funded by the EU network MRTN--CT--2003--504712. The EPSRC funded RB (GR/S10377 \& EP/D067650) and WCKP (EP/D071070). AnM holds an RSE/BP Trust Research Fellowship.

\bibliographystyle{apsrev}
\bibliography{isa_bibliography}

\end{document}